\pdfoutput=1
\documentclass[aps,prb,nofootinbib,twocolumn,floatfix,
groupedaddress,superscriptaddress,reprint,longbibliography]{revtex4-2}

\usepackage{latexsym} 
\usepackage{amsmath} 
\usepackage{dcolumn}
\usepackage{bm} 
\usepackage{graphicx} 
\usepackage{epsfig}
\usepackage{float}
\usepackage[]{graphicx}
\usepackage[]{wrapfig}
\usepackage[]{color}
\usepackage[colorlinks=True]{hyperref}

\begin{document}

\preprint{APS/123-QED}
\title{Theoretical study of high harmonic generation in monolayer NbSe$_2$}

\author{Daniel A. Rehn}
\email{rehnd@lanl.gov}
\affiliation{Computational Physics Division, Los Alamos National Laboratory, Los Alamos, New Mexico 87545, USA}

\author{Towfiq Ahmed}
\affiliation{National Security Directorate, Pacific Northwest National Laboratory, Richland, Washington 99354, USA}

\author{Jinkyoung Yoo}
\affiliation{Center for Integrated Nanotechnologies, Los Alamos National Laboratory, Los Alamos, New Mexico 87545, USA}
    
\author{Rohit Prasankumar}
\affiliation{Center for Integrated Nanotechnologies, Los Alamos National Laboratory, Los Alamos, New Mexico 87545, USA}

\author{Jian-Xin Zhu}
\email{jxzhu@lanl.gov}
\affiliation{Theoretical Division, Los Alamos National Laboratory, Los Alamos, New Mexico 87545, USA}
\affiliation{Center for Integrated Nanotechnologies, Los Alamos National Laboratory, Los Alamos, New Mexico 87545, USA}

\date{\today}

\begin{abstract}
High harmonic generation (HHG) is a powerful probe of electron
dynamics on attosecond to femtosecond timescales and has been
successfully used to detect electronic and structural changes in
solid-state quantum materials, including transition metal
dichalcogenides (TMDs).  Among TMDs, bulk NbSe$_2$ exhibits charge
density wave (CDW) order below 33 K and becomes superconducting below
7.3 K. Monolayer NbSe$_2$ is therefore interesting as a material whose
different structural and electronic properties could be probed via
HHG.  Here, we predict the HHG response of the pristine 2H and CDW
phases of monolayer NbSe$_2$ using real-time time-dependent density
functional theory under the application of a simulated laser pulse
excitation.  We find that due to the lack of inversion symmetry in
both monolayer phases, it is possible to excite even harmonics and
that the even harmonics appear as the transverse components of the
current response under excitations polarized along the zigzag
direction of the monolayer, while odd harmonics arise from the
longitudinal current response in all excitation directions. This
suggests that the even and odd harmonic response can be controlled via
the polarization of the probing field, opening an avenue for
potentially useful applications in opto-electronic devices.
\end{abstract}


\maketitle

\section*{\label{sec:level1}Introduction}
Transition metal dichalcogenides (TMDs) have attracted much attention
recently for their useful and novel physical
properties.~\cite{choi2017recent,akinwande2017review} In bulk form,
most TMDs exist as stacked two-dimensional layers where the atoms
within each layer exhibit strong covalent bonding and layers are held
together by weaker van der Waals interactions. These bulk structures
can be exfoliated down to a single layer (i.e., monolayer), and most
TMD monolayers exhibit interesting properties not found in the
bulk. Many monolayer TMDs exhibit structural polymorphism, in which
the individual layers can exist in multiple structural phases that can
be exploited for a variety of applications.~\cite{manzeli20172d} As
such, monolayer TMDs are currently being investigated for a wide range
of potential applications, such as their use in phase-change
memory,\cite{li2016structural,rehn2018theoretical,wang2017structural}
opto-electronics,\cite{wang2012electronics}
sensors,~\cite{joshi2018review,lee2018two,ping2017recent} straintronic
devices,\cite{akinwande2017review,wei2017straintronics} and more.

Several monolayer and bulk TMDs exhibit charge density wave (CDW)
order and superconductivity at low temperatures, as seen in
NbS$_2$,~\cite{leroux2018traces,van1966superconductivity,yan2019thickness,guillamon2008superconducting,kavcmarvcik2010specific}
NbSe$2$,~\cite{marezio1972crystal,sanchez1995specific,ricco1977fermi,straub1999charge,malliakas2013nb,rossnagel2001fermi,tonjes2001charge,suderow2005pressure,johannes2006fermi,borisenko2009two,calandra2009effect,weber2011extended,soumyanarayanan2013quantum,arguello2014visualizing,arguello2015quasiparticle,leroux2015strong,ugeda2016characterization,silva2016electronic,bischoff2017nanoscale,chen2020strain}
TaS$_2$,~\cite{nagata1992superconductivity,sipos2008mott,navarro2016enhanced,tsen2015structure}
and TaSe$_2$,
~\cite{kumakura1996charge,smith1985band,ryu2018persistent,lian2019coexistence}
making them potentially useful for controlling these quantum
phases with external stimuli, not only in the form of temperature, but
also with other non-thermal external stimuli such as electrostatic
doping~\cite{bischoff2017nanoscale} and strain.~\cite{chen2020strain}

In recent years, high harmonic generation (HHG) has been studied in
the context of both bulk~\cite{kemper_theoretical_2013,mcdonald_theory_2015,luu_high-order_2016,wu_multilevel_2016,ghimire_high-harmonic_2019,yu_high_2019,park_recent_2022,yue_introduction_2022} and monolayer
materials~\cite{liu_high-harmonic_2017,le_breton_high-harmonic_2018,tancogne-dejean_atomic-like_2018,zhang_generating_2018,liu_polarization-resolved_2020,cao_inter-half-cycle_2021,mrudul_high-harmonic_2021,khan2022optical,yue_signatures_2022,jimenez-galan_orbital_2023} with important applications. HHG has
been used as a probe to detect the 2H-to-1T' structural phase change
in monolayer MoTe$_2$ upon electrostatic
doping,~\cite{wang2017structural,zakhidov2020reversible} topological
phase transitions in WTe$_2$,~\cite{sie2019ultrafast} and to study
harmonic orders of up to 13 in monolayer MoS$_2$.~\cite{liu2017high}
In addition, recent experiments have shown the appearance and
tunability of non-integer harmonics in the topological insulator
Bi$_2$Te$_3$.~\cite{schmid2021tunable} Such studies demonstrate the
utility of HHG as a useful attosecond to femtosecond probe of a wide
range of materials that can be used in a variety of applications.

\begin{figure}
  \includegraphics[width=0.40\textwidth]{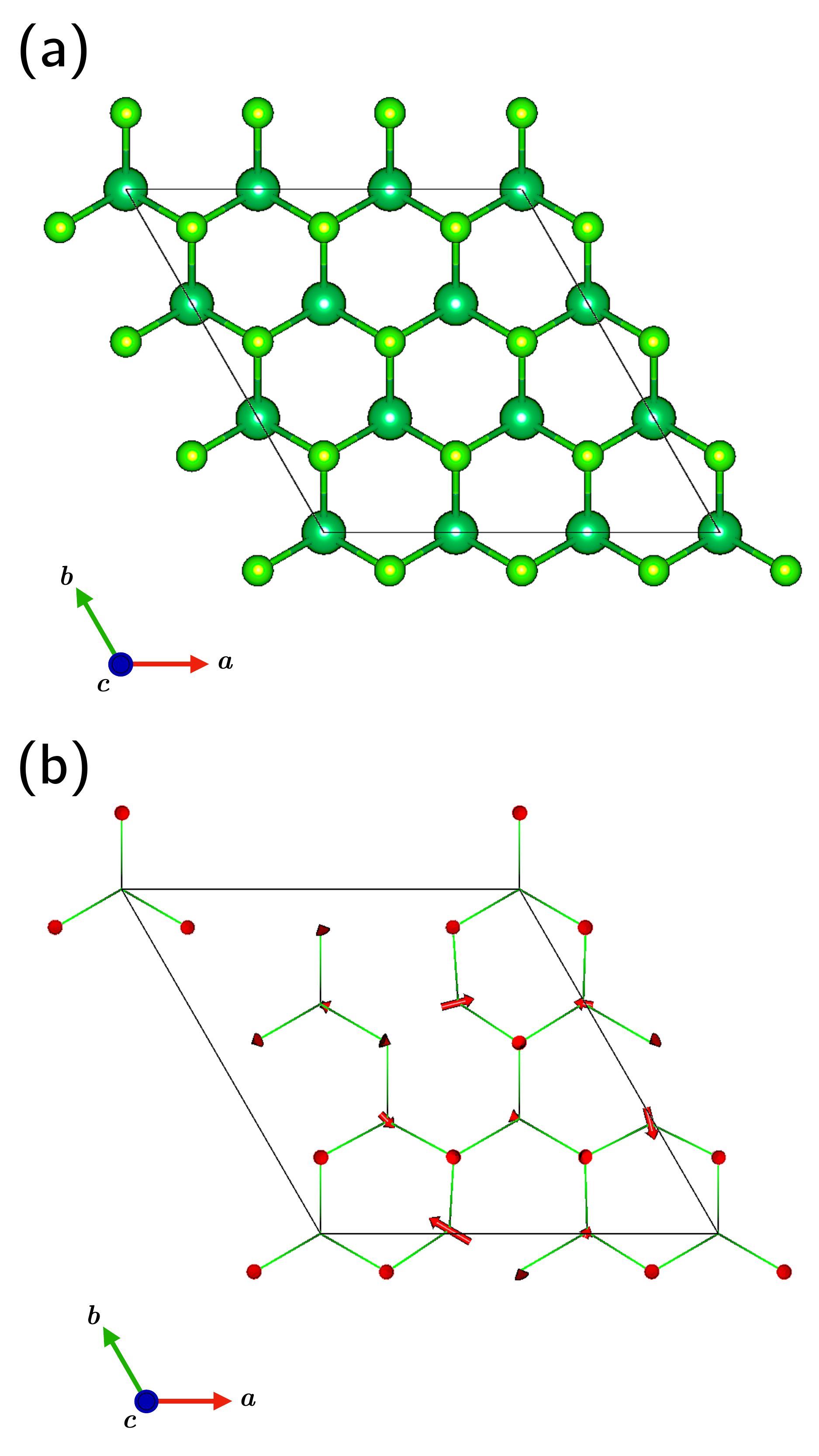}
  \caption{Unit cell of (a) 2H monolayer and (b) a monolayer of the
    CDW phase, with arrows representing enlarged atomic distortion
    directions from the 2H monolayer. Nb atoms are dark green, Se
    atoms are light green. Simulations are performed for both cells
    shown, each of which consists of 27 atoms.}
  \label{fig:1}
\end{figure}

In this work, we study the HHG of monolayer NbSe$_2$ using real-time
time-dependent density functional theory (RT-TDDFT) simulations. By
exciting the monolayer with a simulated terahertz femtosecond laser
pulse, we can predict the current response and HHG spectra under a
variety of pulse orientations and strengths, as the real-time
propagation allows us to probe both the linear and nonlinear
electronic response regimes. Doing so, we find that monolayer NbSe$_2$
exhibits a strong anisotropy in the strength of the transverse current
response when excited along the [100] crystal axis ($x$ or zigzag
direction, oriented along the $\textbf{a}$ lattice constant in
Fig.~\ref{fig:1}) and [010] crystal axis ($y$ or armchair direction,
perpendicular to $\textbf{a}$ in
Fig.~\ref{fig:1}).~\cite{akinwande2017review}
In the case of excitation along the armchair direction, we find that
the even modes are largely suppressed for both the longitudinal
(current parallel to pulse orientation) and transverse (current
perpendicular to pulse orientation) current response.  However, for
laser pulses oriented along the zigzag direction, we find that the
transverse components of the current response are strongly amplified for
excitation strengths of 10$^{11}$ and 10$^{12}$ W/cm$^2$, well above
the linear response regime.

In the case of excitation along the zigzag direction, the longitudinal
HHG spectrum consists primarily of odd-numbered multiples of the
excitation frequency ($n\omega_0$, $n=1,3,5,\dotsc$ with $\omega_0$
the laser pulse frequency). The inversion
symmetry breaking of the 2H monolayer, as compared to its 2H bulk
form, allows in general for even modes ($n\omega_0$, $n=2,4,\dotsc$)
to be present. We find that these even modes appear in the HHG spectra
as the transverse components of the current for pulses oriented along
the zigzag direction. In this sense, the transverse HHG spectra under
excitation along the zigzag direction could be exploited for use in
device applications.  To summarize, the odd-numbered modes of the HHG
response are present in exciations oriented along both the zigzag and
armchair directions, but the even mode response occurs only for
excitations along the zigzag direction, and in that case the even
modes occur only in the transverse direction, so that the
direction-dependence of both the exciting laser pulse and measurement
orientation may be exploited.

In addition to the current response, we investigate the effect of CDW
distortion on the HHG spectrum of monolayer NbSe$_2$. We find that the
CDW distortion, determined from prior experiments, does not lead to an
appreciable change in the HHG spectrum as compared to the pristine 2H
monolayer. This is a result of the small atomic displacements of the
CDW phase as compared to the 2H phase that lead to only small changes
in the the electronic structure of the CDW phase. To understand
whether this would remain the case under larger distortions, we
perform calculations under enlarged distortion conditions. We find
that for these enlarged distortions, noticeable changes arise in the
HHG spectra only when the ground state electronic structure changes
appreciably from the 2H phase.  This indicates the possibility that
for materials exhibiting a larger structural change upon CDW formation
than in NbSe$_2$, the pristine-to-CDW structural change could lead to
distinct changes in the HHG spectra, opening the possibility of
probing CDW distortion on femtosecond timescales.


\section*{Results\label{sec:results}}
\subsection*{Laser pulse excitation}
Using the methodology discussed in the Methods section, we simulate
the application of a femtosecond laser pulse for pulses polarized
along both the zigzag armchair directions with a pulse shape whose
Cartesian component $\alpha=x,y$ is defined as
\begin{equation}
  A_\alpha(t)=-c\frac{E_o}{\omega_o}\cos(\omega_o t)\sin^2\left(\frac{\pi t}{\tau}\right)\Theta(\tau - t),
\label{eq:9}
\end{equation} 
where the electric field can alternatively be defined as: 
\begin{align}
  E_\alpha(t) &= -\frac{1}{c}\frac{\partial A_\alpha(t)}{\partial t} \nonumber \\
  &= \frac{E_o}{\omega_o}\frac{\partial}{\partial t}\left[\cos(\omega_o t)\sin^2\left(\frac{\pi t}{\tau}\right) \Theta(\tau - t)\right ].
  \label{eq:10}
\end{align}

\begin{figure*}[!ht]
  \includegraphics[width=\textwidth]{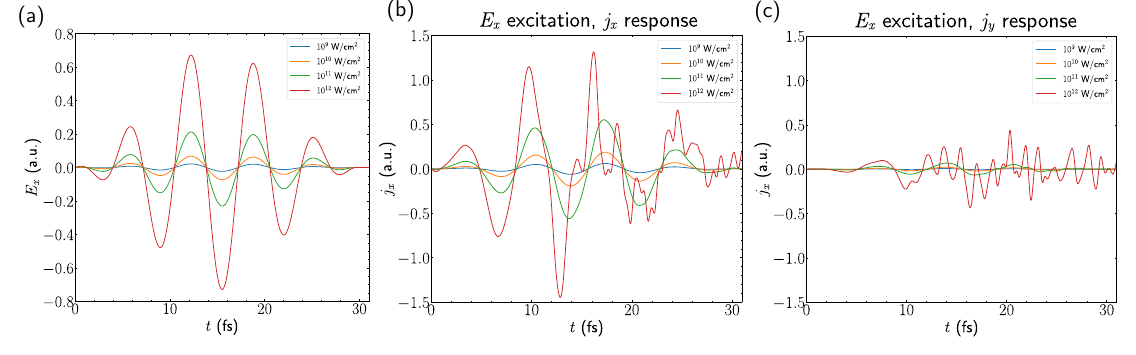}
  \caption{Real-time excitation pulse and current response for
    excitations along the zigzag ($E_x$) direction, applied to the
    pristine 2H monolayer. (a) $E_x(t)$ for different pulse
    intensities, (b) Longitudinal current response $j_x(t)$ for different
    pulse intensities and (c) transverse current response ($j_y$) for
    different intensities.}
  \label{fig:2}
\end{figure*}

In our calculations, we use a pulse-width of $\tau$ = 30 fs and photon
energy $\omega$ = 0.6 eV $\approx$ 145 THz. The intensity of the pulse
is related to the amplitude by $I$ $\approx$ c$E^2_o$/8$\pi$. This
equality only holds for infinite unmodulated wave trains (not pulses),
but the approximation can be used to determine a corresponding value
for $E_0$ in Eqs.~\ref{eq:9}--\ref{eq:10}, as described in
Ref.~\onlinecite{Yabana_2}.
The external field $E_\alpha(t)$ pulse shapes for four different
intensities, $10^9$, $10^{10}$, $10^{11}$, and $10^{12}$ W/cm$^2$, are
shown in Fig.~\ref{fig:2}a.

Calculations for both CDW and non-CDW structures were performed for a
31 fs time-propagation using $\Delta t$ = 0.08 $\hbar/E_H$ atomic
units, or 1.935 attoseconds as the time step. This timestep was
determined to give the same HHG spectrum as simulations run with
$\Delta t = 0.02\;\hbar/E_h$ (see Supp. Material Sec. 4). For
time-propagation, we used the enforced time-reversal symmetry
propagator.~\cite{gomez2018propagators} We also point out that there
is a subtle issue associated with how to accurately extract HHG
information from RT-TDDFT simulations that use periodic boundary
conditions. The issue that arises is that for long enough simulation
times, the current can start to propagate through periodic repeats of
the cell, which is unphysical and would not be seen
experimentally. This issue so far has been addressed in two different
ways. One is to use a multiscale modeling approach in which the
microscopic current calculated in the RT-TDDFT approach is used as
input to a macroscopic calculation that solves Maxwell's
equations.~\cite{yabana2012time,sato2014maxwell} The other approach is
to run a normal RT-TDDFT simulation that is short enough to avoid the
issue of the current interfering with its periodic repeats as a result
of the application of periodic boundary conditions and a relatively
small cell size.~\cite{floss2018ab} Note that a related issue is
ultrashort dephasing times.~\cite{brown_real-space_2024} Both
approaches have been shown to produce useful qualitative
information. Therefore, in this study we take the second approach and
evolve the TDKS equations for only 1 fs after the 30 fs laser pulse.
We extensively tested this issue and found that the results do not
change much qualitatively for any total simulation time from 30-35 fs,
as discussed in Supp. Material Sec. 6. We also point out that
experimentally, HHG spectra are usually collected only for the
duration of the excitation pulse, in agreement with our computational
approach.

\begin{figure*}[!ht]
  \includegraphics[width=\textwidth]{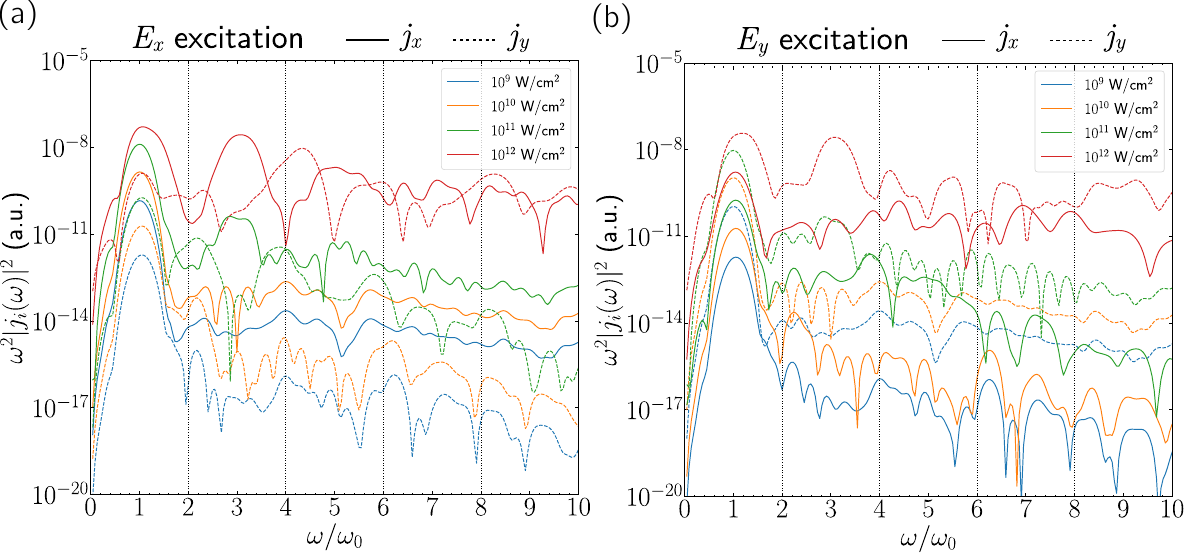}
  \caption{HHG spectra for different excitation strengths in the 2H
    monolayer. (a) $E_x$ excitation with $j_x$ (solid) and $j_y$
    (dashed) response. (b) $E_y$ excitation with $j_x$ (solid)
    response and $j_y$ (dashed) response.}
  \label{fig:3}
\end{figure*}

\subsection*{Transverse HHG in the 2H monolayer}
With application of the electric field pulse described in Methods, we
compute the current response throughout the pulse duration (30 fs) and
for 1 fs after (31 fs total).  As an example, we plot the longitudinal
($j_x$) and transverse ($j_y$) currents under excitation in the $E_x$
direction in Fig.~\ref{fig:2}b and Fig~\ref{fig:2}c,
respectively. This shows that the transverse components are much
smaller in magnitude than the longitudinal current, though they are
not negligible for the higher pulse strengths. In addition, we see a
clear nonlinear response of the current after the first $\sim 10$ fs
for the $10^{12}$ W/cm$^2$ case, and although it is not as obvious for
$10^{11}$ W/cm$^2$, the nonlinear response is also present. By
nonlinear, we mean that the current response does not respond linearly
with the driving field and higher frequency oscillations in the
current appear.  The nonlinear transverse currents arise from the
nonzero Berry curvature of bands.~\cite{avetissian_high_2020}

In Fig.~\ref{fig:3} we plot the HHG spectra (calculated via
Eq.~\ref{eqn:eqn_06a}) for the $E_x$ (panel a) and $E_y$ (panel b)
excitations of the pristine 2H monolayer. The blue, orange, green, and
red lines correspond to pulse intensities of $10^n$ W/cm$^2$,
$n=9,10,11,12$, respectively. The solid lines are the $j_x$ current
while the dashed lines are the $j_y$ current.  In panel a, the $j_x$
current is the dominant contribution to the total current, being the
longitudinal current response, whereas $j_y$ is the dominant contribution
for the $E_y$ excitation, as in that case it is the longitudinal current. In
addition, we see that in the $10^9$ and $10^{10}$ W/cm$^2$ cases, the
only prominent response is at the pulse freuqency
$\omega=\omega_0$. This is a clear indication that these pulse
intensities are not strong enough to induce a nonlinear current
response that is strong enough to observe a clear HHG signal.  In
contrast, we see that for pulse intensities of $10^{11}$ and $10^{12}$
W/cm$^2$, we begin to see higher harmonics appear in the longitudinal
current response for both $E_x$ excitation ($j_x$ response) and $E_y$
excitation ($j_y$ response). This indicates that these pulse
intensities are strong enough to witness HHG.

In addition, we see that for the zigzag ($E_x$) excitation direction,
the $10^{11}$ and $10^{12}$ W/cm$^2$ intensities give rise to a large
$j_y$ contribution, i.e., the transverse current response.  In fact, for
$10^{11}$ W/cm$^2$, the second harmonic is nearly as strong as the
fifth harmonic in $j_x$, and in the $10^{12}$ W/cm$^2$ case, the
fourth harmonic is nearly as strong as the third harmonic in
$j_x$. On the other hand, excitation in the armchair ($E_y$)
direction does not lead to an appreciable transverse ($j_x$) response in any
of the cases. For monolayers with space group $P\bar6m2$, no inversion
symmetry is present, so it is possible to observe even harmonics in
the HHG spectrum. This is in contrast to the bulk 2H phase, where an
inversion center is present in the van der Waals gap between NbSe$_2$
layers.

\begin{figure*}[!ht]
  \includegraphics[width=\textwidth]{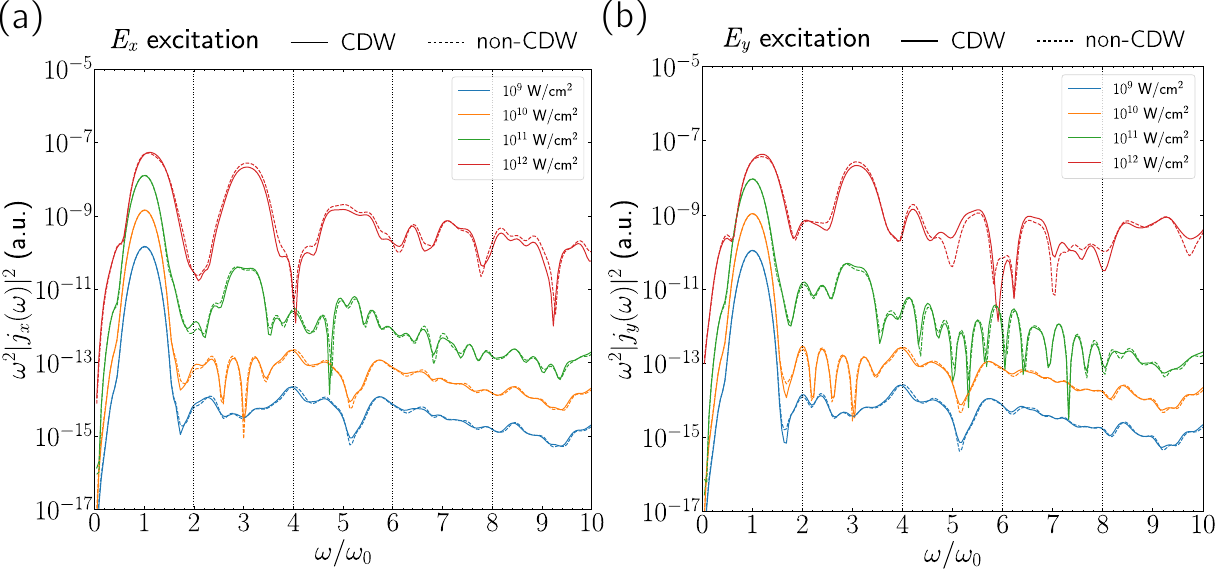}
  \caption{Comparison of the longitudinal current HHG spectra for the CDW
    (solid lines) and 2H (dashed lines) cells at different pulse
    intensities. (a) $E_x$ excitation with $j_x$ response. (b) $E_y$
    excitation with $j_y$ response. Only the longitudinal components are
    included in order to reduce clutter; the corresponding transverse
    components can be found in Supp. Material Sec. 7.}
  \label{fig:4}
\end{figure*}

One question is why we do not see a strong even harmonic response for
excitation along the $E_y$ direction. We first note that in fact, the
even harmonics in $j_y$ are not entirely absent in
Fig.~\ref{fig:3}b. For example, we do see the 2nd harmonic in $j_y$
for $10^{11}$ W/cm$^2$, and we see to some degree both the 2nd and 4th
harmonics for 10$^{12}$ W/cm$^2$.  Nonetheless, the response is not
strong. However, the symmetry considerations can only tell us whether
a response is allowed or forbidden, and when allowed, the magnitude of
response depends on the details of the underlying electronic
structure, which must be computed.

The findings above indicate some potential opto-electronic device
applications that we now discuss.  Because even harmonics are only
found for propagation along the [010] (or another symmetry-equivalent)
direction, and because these even harmonics are only substantially
excited for excitation pulses oriented along the [100] direction, it
may be possible to use the even harmonic response as a femtosecond
switch to pass information along.  For example, given a large enough
single-crystal monolayer, it would be possible to apply laser pulses
oriented along different directions and simultaneously collect the
response. Similar to bits that are either 0 or 1, the appearance of an
even harmonic within some femtosecond or shorter window could
represent a 1, and the non-appearance could represent a 0. If the
pulse intensity were also used as an input variable, it could be
possible to go beyond binary channels and pass information along 0 (no
even harmonic), 1 (2nd harmonic), and 2 (fourth harmonic), etc. By
constructing rules for how one type of response (appearance of a
particular harmonic) triggers subsequent excitations, a complex
network of rules could be constructed to filter information on
femtosecond timescales.

\subsection*{HHG in the CDW monolayer}
Having explained the prediction of the transverse HHG response for even
harmonics in the pristine 2H monolayer, we now turn to the question of
how the HHG response of the CDW phase compares with that of the 2H
monolayer. Using the same computational approach, we make comparisons
of only the longitudinal current response in Fig.~\ref{fig:4} (the
transverse current responses are shown in Supp. Material Sec. 7). We include
only the longitudinal current in order to more easily see the
comparisons and reduce clutter in the figure.  Clearly, the HHG
spectra look extremely similar for both the pristine (dashed lines)
and CDW (solid lines) phases. Although we do see some modest
differences, particularly for the 10$^{12}$ W/cm$^2$ case in panel b,
these differences are likely not prominent enough to be useful for
device applications.  These findings are also true for the transverse HHG
responses shown in Supp. Material Sec. 7, where again there are some
noticeable differences in the $10^{12}$ W/cm$^2$ case, though they may
not be large enough to clearly differentiate phases experimentally.

One interesting question that arises is why the CDW distortion does
not lead to a significant change in the HHG properties. We point out
that since the CDW phase also lacks inversion symmetry, the even
harmonics may be present.  For NbSe$_2$, the atomic displacements in
moving from the pristine to CDW phase are quite small and comparisons
of the ground state electronic density of states (DOS) show that the
distortions are not enough to significantly change the DOS.  Although
it is not \emph{a-priori} clear that close agreement of the ground
state DOS for the two phases is sufficient to say that the HHG spectra
will be close, we find that this is borne out in our numerical
simulations. However, it will be important to investigate this
experimentally, as well.

Although we do not find significant differences in the HHG spectra for
the CDW and pristine phases of NbSe$_2$, the case may be different for
other materials with larger CDW structural changes.  Therefore, one
interesting question is whether a theoretically enlarged CDW
distortion of the CDW phase of NbSe$_2$ could lead to differences in
the HHG spectra.  To explore this, we employ the following technique
to generate CDW structures with enlarged distortions.  Denoting the
positions of the atoms in the pristine 9 f.u. 2H cell as
$\textbf{r}_i^{\rm 2H}$ $(i=1,2,\dotsc 27)$ and the positions of atoms
in the 9 f.u. CDW cell as $\textbf{r}_i^{\rm CDW}$
($i=1,2,\dotsc,27$), we define a new set of atomic positions
\begin{equation}
  \textbf{r}_i^\alpha = \textbf{r}_i^{\rm CDW} + \alpha \big( \textbf{r}_i^{\rm CDW} - \textbf{r}_i^{\rm 2H}\big).
  \label{eq:ria}
\end{equation}

\begin{figure*}[!ht]
  \includegraphics[width=\textwidth]{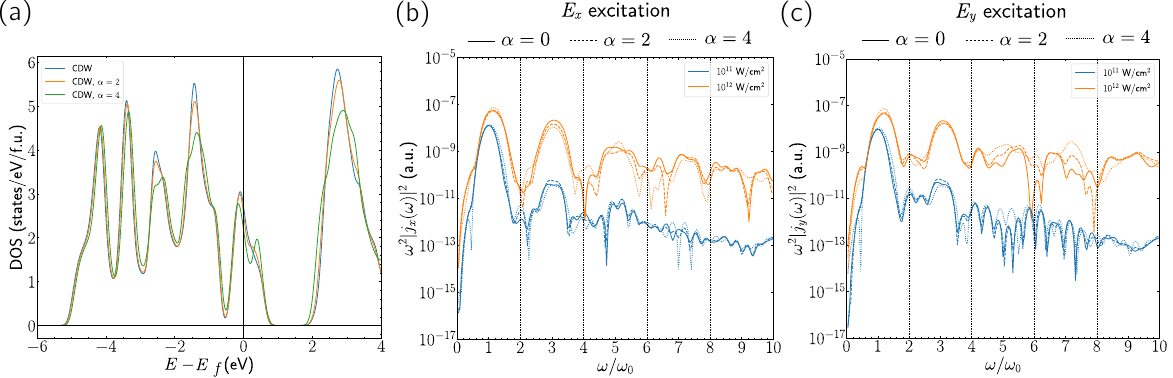}
  \caption{(a) Ground state DOS for the CDW ($\alpha=0$) and
    enlarged distortion ($\alpha=2,4$) structures. (b) Longitudinal HHG
    spectra for $E_x$ excitation and $j_x$ response and (c) Longitudinal HHG
    response for $E_y$ excitation and $j_y$ response, for $10^{11}$
    (blue) and $10^{12}$ (orange) W/cm$^2$ pulse intensities for the
    three values of $\alpha$.}
  \label{fig:5}
\end{figure*}

This definition allows us to use different values of $\alpha$ to
enhance the structural distortion, with $\alpha=0$ corresponding to
the experimental CDW structure. In Fig.~\ref{fig:5} we present
comparisons of the longitudinal HHG spectra for $E_x$ and $E_y$ excitations
for $\alpha=0,2,4$ in panels b and c, respectively.  For the case of
$\alpha=4$, the distortion is large enough to start to see noticeable
qualitative changes in the HHG spectra. This is also true for the transverse
HHG spectra shown in Supp. Material Sec. 8, where we see perhaps even
larger qualitative changes to the HHG spectra.

The case of $\alpha=4$ also produces a significant change in the
ground state DOS as compared to the $\alpha=0$ case, as seen in
Fig~\ref{fig:5}a. Therefore, from the basis of these calculations, one
indicator for whether we might expect to see qualitative changes in
the HHG spectra of CDW and non-CDW phases of other materials is if
their ground state DOS are significantly different. Of course, CDW
distortions that lead to changes in space group, particularly in cases
where an inversion symmetry is broken, could also be expected to lead
to large qualitative differences in the HHG spectra. In this case,
both $\alpha=2,4$ have space group $P\bar6$ (\#174), which is a lower
symmetry than the CDW phase and where inversion symmetry is again
absent.

While we do not find it useful to discuss in detail the myriad of
changes in the HHG signatures of the structures with enlarged
distortions (since those structures would not be realized easily in
experiments), the investigation shows that at least theoretically, CDW
distortions can lead to distinct HHG spectra that could be detected
experimentally and potentially exploited for device
applications. Other materials that exhibit large structural
distortions could therefore be interesting to explore both in
simulations and experiments.

\section*{Discussion \label{sec:discussion}}
We have presented results from a computational study of laser pulse
excitations of monolayer NbSe$_2$ in its pristine 2H and CDW
phases. Different pulse orientations and strengths lead to the
following predictions: (1) the longitudinal current response for
excitations along both the zigzag and armchair directions of NbSe$_2$
leads to an HHG signature that is dominated by odd harmonics for
excitation strengths of $\sim 10^{11}$ W/cm$^2$ and higher; (2) the
transverse HHG response is extremely pronounced in case of excitations along
the zigzag direction and appear predominantly as even harmonics,
whereas excitations along the armchair direction do not lead to such a
prominent appearance of even harmonics in neither the longitudinal nor
transverse HHG components; (3) the HHG response of the experimentally
measured CDW phase does not reveal any significant changes from the
pristine 2H phase, due to the small atomic displacements, however (4)
enlarging the CDW distortion leads to significant qualitative changes
in the HHG spectra, indicating that other materials with larger CDW
structural distortions than those of NbSe$_2$ may lead to distinct
differences in HHG spectra.

The findings summarized here indicate some potential ways to exploit
the findings for useful opto-electronic device applications. In
particular, since the even harmonics only appear strongly as the
transverse part of the current response for excitations along the
zigzag direction, it may be possible to use the detection of even
harmonics as a binary, ternary, or higher order switch to pass along
information. The potential benefit for opto-electronic applications is
that devices could be made atomically-thin and operate on femtosecond
or faster timescales. Experimental investigation of the predictions
presented here would shed light on the feasibility of using NbSe$_2$
or other materials for this, and other, device applications.

\section*{Methods\label{sec:methods}}
The real-time formulation of time-dependent density functional
theory~\cite{runge1984density} is carried out in a time-dependent
Kohn-Sham (KS) scheme in which the KS wavefunctions
$\psi_i(\textbf{r}, t)$ are evolved via~\cite{ullrich2011time}
\begin{equation}
  i\hbar\frac{\partial}{\partial t} \psi_i(\textbf{r},t)
   = \hat{H}_\mathrm{KS} \psi_i(\textbf{r},t),
   \label{eq:1}
\end{equation}
where $i$ is the band index and
\begin{align}
  \hat{H}_\mathrm{KS} = \frac{\textbf{p}^2}{2 m} &+
  v_\mathrm{ion}(\mathbf{r}) +
  e^2\int d\textbf{r}' \frac{n(\textbf{r}',t)}{|\textbf{r} - \textbf{r}'|}
  \nonumber \\
  &+ v_{xc}[n(\textbf{r},t)] + e \textbf{E}(t) \cdot \textbf{r}.
  \label{eq:2}
\end{align}
and $\mathbf{p}=-i\hbar\nabla$, $m$ is the electron mass, and
$v_\mathrm{ion}(\mathbf{r})$ is the potential of the ions, assumed
here to be static and treated using pseudopotentials as discussed
below. The third term is the Hartree Coulomb interaction, the fourth
term $v_{xc}[n]$ is the exchange-correlation potential, and the
electron density is
\begin{equation}
  n(\textbf{r},t) = \sum_i |\psi_i(\textbf{r},t)|^2.
\end{equation}
$\textbf{E}(t)$ is the time dependent electric field. The
time-dependent external interaction potential term $e \textbf{E}(t)
\cdot \bf {r}$ breaks the translational symmetry of periodic
systems. This can be addressed by adopting the velocity gauge, where a
vector field, defined as~\cite{yabana2012time}
\begin{equation}
  \textbf{A}(t) = -c\int^t \textbf{E}(t') dt',
  \label{eq:3}
\end{equation} 
is used to gauge-transform the KS wave functions as~\cite{yabana2006real}
\begin{equation}
  \psi(\textbf{r},t) \rightarrow \exp\left[\frac{ie}{\hbar c} \textbf{A}(t)
    \cdot \textbf{r}\right] \psi(\textbf{r},t).
  \label{eq:4}
\end{equation}
The velocity gauge TDKS Hamiltonian now takes the form
\begin{align}
 \hat{H}^\mathrm{RT}_\mathrm{KS} &= \frac{1}{2m}\left[\textbf{p} +
   \frac{e}{c}\textbf{A}(t)\right]^2 + v_\mathrm{ion}(\mathbf{r})
 \nonumber \\ &+ e^2\int d\textbf{r}'
 \frac{n(\textbf{r},t)}{|\textbf{r} - \textbf{r}'|} +
 v_{xc}[n(\textbf{r},t)].
 \label{eq:5}
\end{align}
The coupling to the external field is now incorporated in the kinetic
energy term, and consequently the translational symmetry of the
Hamiltonian is restored. The time-dependent KS orbitals are now
evolved using
\begin{equation}
  i\hbar\frac{\partial}{\partial t}
  \psi_i(\textbf{r},t)=\hat{H}^\mathrm{RT}_\mathrm{KS}
  \psi_i(\textbf{r},t).
  \label{eq:6}
\end{equation}
In RT-TDDFT a time evolution propagator is needed to evolve
Eq.~\ref{eq:6} in time. Several options for doing so for solids
have been discussed in the
literature,~\cite{castro2004propagators,gomez2018propagators,schleife2012plane,rehn2019ode}
with a detailed discussion found in
Ref.~\onlinecite{castro2004propagators}.

By propagating the KS wave functions, the time-dependent current
density
\begin{eqnarray}
  \textbf{j}(t) = -\frac{i}{2\Omega}\int_{\Omega} d\textbf{r} \sum_i \left[ \psi^*_i(\textbf{r},t)\nabla \psi_i(\textbf{r},t) - \text{c.c.}\right]
\label{eqn:eqn_05a}
\end{eqnarray} 
can be obtained, whose power spectrum gives the HHG response
\begin{eqnarray}
\label{eqn:eqn_06a}
\textbf{HHG}(\omega)=\omega^2\left|\int^T_0 \textbf{j}(t)\exp(-i\omega t)dt\right|^2.
\end{eqnarray} 
Here $\Omega$ is the spatial volume of the unit cell and $T$ in the
Fourier transformation is the total propagation time for our
simulation.

The velocity gauge formalism for RT-TDDFT has been implemented in
several codes, including TDAP~\cite{tdap},
RT-SIESTA,~\cite{velocity_gauge} Elk,~\cite{krieger2015laser,elk}
OCTOPUS,~\cite{octopus} and SALMON,~\cite{salmon_1} among others. We
used SALMON version 2.0.2 for all calculations presented in this
article. SALMON is a real space code with options to apply periodic
boundary conditions, so it is well-suited for studying monolayer
NbSe$_2$.

All calculations are performed for fixed atomic positions, since the
maximum simulation times are on the order of tens of femtoseconds,
much faster than the vibrational timescales of the nuclei. In
addition, all calculations are performed under the adiabatic
approximation within TDDFT, which is to say that we neglect the
history dependence in the evaluation of the exchange-correlation
functional at each time and instead evaluate $v_{xc}[n]$ using the
instantaneous $n(\textbf{r}, t)$.  Within the adiabatic approximation
we employ the local density approximation (LDA)~\cite{yabana1996time}
for all calculations.  SALMON makes use of pseudopotentials and for
all calculations we use Trouiller-Martin type norm-conserving LDA-FHI
pseudopotentials~\cite{PP_1,PP_2} for the Nb and Se atoms, where 5
($4d^45s^1$) and 6 ($4s^24p^4$) electrons were treated in the valence,
respectively.  We verified the accuracy of these pseudopotentials for
reproducing the ground state electronic structure of monolayer
NbSe$_2$ by comparing the ground state density of states (DOS) to
highly accurate all-electron calculations performed in the Elk
code~\cite{elk} version 7.0.12 (see Supp. Material Sec. 3 for
details). The density of states we calculate is in good agreement with
X-ray photoemission experiments.~\cite{ricco_density_1976} Also note
that we do not include spin orbit coupling (SOC) effects in this work,
as this capability is not available in the code at present. SOC
effects were studied recently in the context of the CDW
phase.~\cite{nakata_anisotropic_2018}

For all calculations we used hexagonal unit cells for both the
pristine 2H monolayer and CDW monolayer that consist of 9 formula
units (f.u.) of NbSe$_2$ (27 atoms in total). The pristine monolayer
and CDW space groups are both $P\bar6m2$ (\#187) within the standard
tolerances of the symmetry software used.~\cite{curtarolo2012aflow}
The monolayer structures are found by isolating a single layer of the
bulk 2H phase, space group $P6_3/mmc$ (\#194), and CDW phase, space
group $P6_3/m$ (\#176). For the CDW phase, the 9 f.u. cell corresponds
to the primitive cell, while for the 2H phase this corresponds to a
$3\times3\times1$ supercell (see Fig.~\ref{fig:1}). In order to
isolate the effects of CDW distortion in our calculations, we chose to
use the same lattice constants for both the pristine and CDW phases,
where we use the CDW phase lattice constants determined by Malliakas
et al. (Ref.~\onlinecite{malliakas2013nb}) measured at 15 K, which is
$|a|=|b|=10.3749$ {\AA}. This corresponds to a primitive 2H unit cell
size of $|a|=|b|=3.4583$ {\AA}, which is slightly expanded from the 2H
lattice constants measured at room temperature,
$|a|=|b|=3.442$--$3.446$ {\AA}, as determined by various
sources.~\cite{kadijk1971polymorphism,malliakas2013nb} This
corresponds to an axial strain of the pristine cell of less than
0.5\%, not accounting for differences arising from thermal
expansion. As NbSe$_2$ is metallic, the use of the bulk experimental
lattice constants, which will in general not correspond exactly to the
equilibrium (zero temperature) lattice constants of the monolayer
predicted within the LDA, is not expected to lead to a significant
qualitative change in the RT-TDDFT results. This situation would be
different for semiconducting and insulating systems, in which small
strains can lead to important qualitative changes that could strongly
influence the current response, e.g., stemming from changes from a
direct to indirect band gap.
Because the monolayer lattice constants are not as well characterized
and depend on the choice of substrate and growth or exfoliation
method, we chose to use the experimentally measured bulk CDW lattice
constants for all monolayer calculations.

All RT-TDDFT calculations require an initial calculation of the ground
state as an initial state used to start the real time propagation with
the laser pulse described above. Because SALMON is a real space code
and we use periodic boundary conditions, we must ensure that both the
ground state and the HHG spectra are converged with respect to the
real-space grid size and k-mesh.  For the ground state, we determined
that a grid density $n_r= 2$ pts/bohr is necessary to ensure that the
wave functions are adequately represented on the grid (see
Supp. Material Sec. 2 for details). Note that accurate ground state
calculations required some adjustments in the mixing parameters of the
Broyden scheme~\cite{broyden1965class} (see Supp. Material Sec. 1 for
details), and the self consistency cycle was stopped after
$\big(\int_\Omega|n_i(\textbf{r})-n_{i-1}(\textbf{r})|d^3r\big)/N_{\rm
  el} < 10^{-10}$, where $n_j(\textbf{r})$ is the charge density
computed for the $j$th step in the selfconsistency cycle and $N_{\rm
  el}$ is the total number of valence states in the cell.

To determine the k-mesh size needed for calculations, we first ensured
that the ground state density of states was converged with respect to
k-mesh size and then performed convergence studies of the HHG spectra
with respect to k-mesh size. We found that for the 9 f.u. cells a
k-mesh of size $8\times8\times1$ is needed to ensure the HHG spectra
are well converged (see Supp. Material Sec. 5 for details). Also note
that no symmetry reductions of the k-mesh are used in calculations, so
as not to enforce unwanted symmetries in the time-dependent Kohn-Sham
wave functions.

\section*{Acknowledgement}
We thank Christos Maliakas for graciously providing us with the CDW
phase crystal structure, as determined from the XRD measurements in
Ref.~\onlinecite{malliakas2013nb}.

This work was carried out under the auspices of the U.S. Department of
Energy (DOE) National Nuclear Security Administration under Contract
No. 89233218CNA000001 and was supported by the LANL LDRD Program under
the Project No. 20190026DR.  We acknowledge the support by the
Institutional Computing Program at LANL and NERSC, via the Center for
Integrated Nanotechnologies, a DOE BES user facility, for
computational resources.

\section*{Data Availability}
All data supporting the findings of this study are available from the corresponding author upon a reasonable request.
 
\section*{Author Contributions}
D.R performed computations and D.R., T.A., and J.-X.Z. analyzed the data. J.Y. and R.P. participated in the discussion of results. J.-X.Z. designed the project and led the investigations, D.R. and T.A. designed the computational approaches. All authors contributed to the writing of the manuscript. Correspondence should be addressed to D.R. (rehnd@lanl.gov) and J.-X.Z. (jxzhu@lanl.gov).
 
\section*{Competing Interests}
The authors declare no competing interests.

\bibliography{refs}{}

\end{document}